\documentclass{aastex631}
\pdfoutput=1
\usepackage{amsmath}
\usepackage[T1]{fontenc}
\usepackage[figure,figure*]{hypcap}
\usepackage{natbib}
\usepackage{hyperref}

\shorttitle{ZTF Forced-Photometry Service}
\shortauthors{Masci, Laher, et al.}

\begin{document}

\title{A New Forced Photometry Service for the Zwicky Transient Facility}

\correspondingauthor{Frank J. Masci}
\email{fmasci@caltech.edu}

\author[0000-0002-8532-9395]{Frank J. Masci}
\author[0000-0003-2451-5482]{Russ R. Laher}
\author[0000-0001-7648-4142]{Benjamin Rusholme}
\author[0000-0003-4401-0430]{David Shupe}
\author[0000-0002-5158-243X]{Roberta Paladini}
\author[0000-0001-5668-3507]{Steve Groom}
\author[0000-0002-9998-6732]{Avery Wold}
\affiliation{IPAC, Mail Code 100-22, Caltech, 1200 E. California Blvd.,
             Pasadena, CA 91125, U.S.A.}

\author[0000-0001-9515-478X]{Adam A. Miller}
\affiliation{Department of Physics and Astronomy, Northwestern University,
             2145 Sheridan Road, Evanston, IL 60208, USA}
\affiliation{Center for Interdisciplinary Exploration and Research in
             Astrophysics (CIERA), 1800 Sherman Ave., Evanston, IL 60201, USA}

\author[0000-0003-0228-6594]{Andrew Drake}
\affiliation{Division of Physics, Mathematics and Astronomy,
             California Institute of Technology, Pasadena, CA 91125, USA}

\begin{abstract}

We describe the Zwicky Transient Facility (ZTF) Forced Photometry Service (ZFPS)
as developed and maintained by the ZTF Science Data System Team at IPAC/Caltech.
The service is open for public use following a subscription. The ZFPS has been
operational since early 2020 and has been used to generate publication quality
lightcurves for a myriad of science programs. The ZFPS has been recently
upgraded to allow users to request forced-photometry lightcurves for up to
1500 sky positions per request in a single web-application submission.
The underlying software has been recoded to take advantage of a parallel
processing architecture with the most compute-intensive component rewritten
in C and optimized for the available hardware. The ZTF
processing cluster consists of 66 compute nodes, each hosting at least 16
physical cores. The compute nodes are generally idle following nightly
real-time processing of the ZTF survey data and when other {\it ad hoc}
processing tasks have been completed. The ZFPS and associated infrastructure
at IPAC/Caltech therefore enable thousands of forced-photometry lightcurves
to be generated along with a wealth of quality metrics to facilitate analyses
and filtering of bad quality data prior to scientific use. 

\end{abstract}


\vspace{0.5cm}
\section{Introduction}\label{sec:intro}

The Zwicky Transient Facility (ZTF) is an optical time-domain survey that
commenced operations in March 2018. ZTF uses the Palomar 48-inch Schmidt
telescope and a custom-built wide-field camera providing a 47 deg$^2$ field
of view and 8 sec readout time, and can scan the entire northern visible
sky at rates of $\sim$3760 square degrees/hour to median depths of $g\sim20.8$
and $r\sim20.6$ mag (AB, 5$\sigma$ in 30 sec). The design and implementation of
the camera and observing system are described in \citet{Bellm:18:ZTF}.
The ZTF Data System at IPAC/Caltech provides near-real-time reduction of the
image data to identify moving and varying objects, and produces astrometrically
and photometrically calibrated source catalogs as the survey proceeds
\citep{Masci:18:ZTF}. ZTF's science objectives were presented in
\citet{Graham:18:ZTF}.

The single-exposure source catalogs generated by the automated processing
pipeline assume a fixed detection threshold along
with filtering of other quality criteria following PSF-fit photometry
\citep[for details, see][]{Masci:18:ZTF}. A consequence of this
fixed-threshold approach is that only measurements of a source that
satisfied the detection threshold (nominally 5$\sigma$) across a span
of exposures are saved and catalogued. Sources that vary and intermittently
fall below the threshold are missed, causing individual lightcurves
constructed from the standard pipeline products to be highly incomplete
and irregular as a function of observation epoch. Furthermore, the
absence of measurements below the detection limit prevents single-epoch
upper limits on a source's location to be determined. A forced photometry
service, customized for ZTF (the ZFPS) was implemented to circumvent
these limitations.

\begin{figure*}
\centering
\includegraphics[scale=0.3]{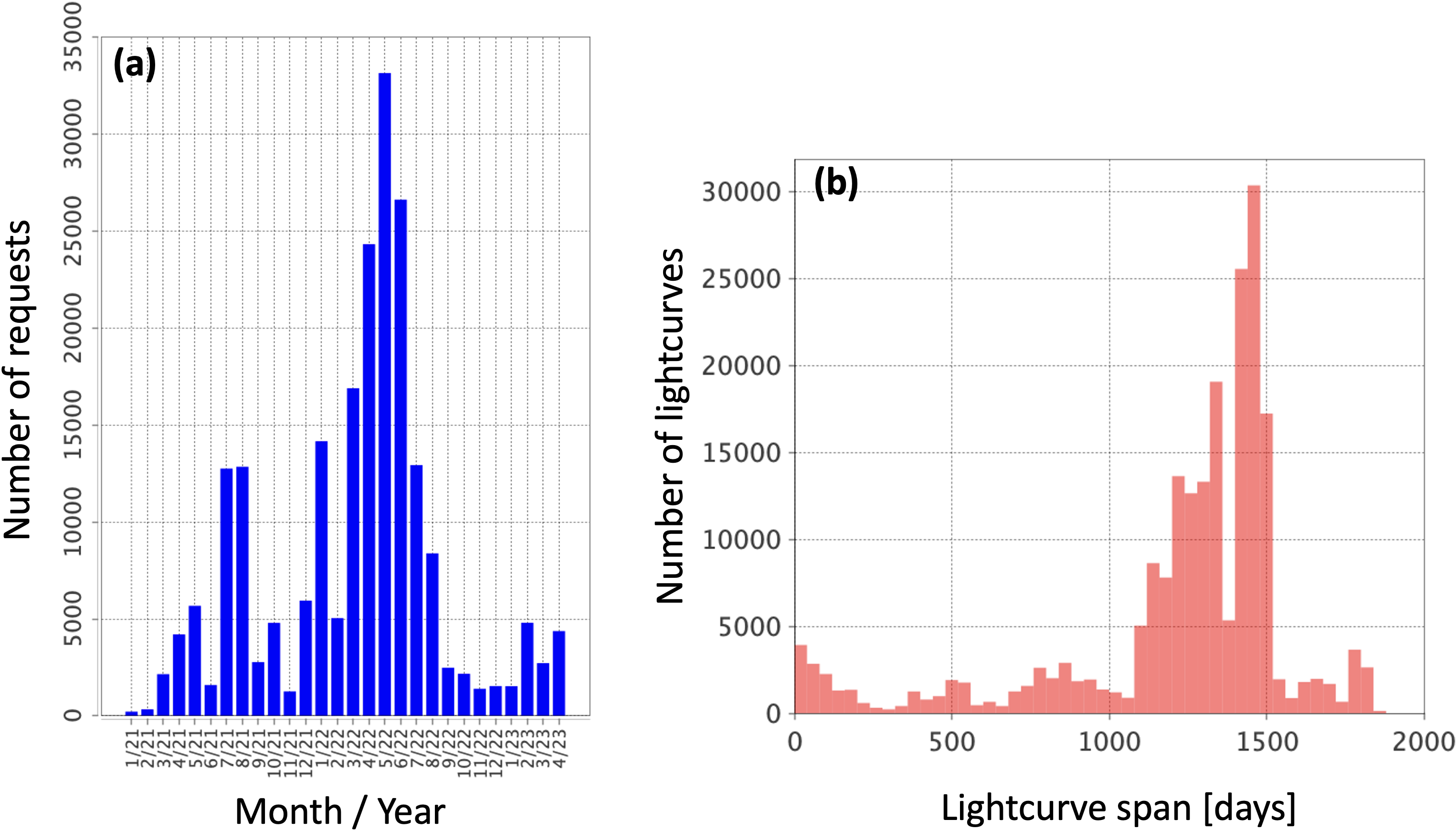}
\caption{(a) Number of forced photometry requests submitted versus calendar
         date since January 1, 2021. (b) Number of forced photometry lightcurves
         generated as a function of lightcurve time-span, also from requests 
         submitted since January 1, 2021.}
\label{fig:zfpshistos}
\end{figure*}

An early implementation of the ZFPS was based on accepting single source
positions with timespans per request\footnote{https://irsa.ipac.caltech.edu/data/ZTF/docs/ztf\_forced\_photometry.pdf}.
Given the prodigious increase in the number of transient and variable
source candidates from the ZTF survey requiring detailed lightcurves
to facilitate classification, there was a demand to improve the processing
throughput of the ZFPS. The new service can handle over 100 times more
individual source requests than the initial version, where users can submit
requests in a batch-like manner. Figure~\ref{fig:zfpshistos}a summarizes the
usage statistics of the service so far, going back to when requests were first
logged into a database (January 2021). Figure~\ref{fig:zfpshistos}b shows the
distribution of lightcurve spans over all requests. The peak at $\sim1450$ days
(approaching fours years) is due to requests to support studies of variability
in Active Galactic Nuclei
\citep[AGN; e.g.,][]{Demianenko:21:ZTF,Demianenko:22:ZTF}.
At this time, there are 256 subscribed users.

This write-up is primarily a user-guide to the new ZFPS. Section~\ref{sec:submission}
describes how to submit forced photometry requests to operate
specifically on difference images stored in the ZTF archive hosted by the
NASA/IPAC Infrared Science Archive (IRSA)\footnote{https://irsa.ipac.caltech.edu/Missions/ztf.html}.
By ``difference images'', we mean images constructed by subtracting a reference
image (co-add) constructed from early ZTF survey image data from
single-epoch science images. Use of difference images for forced photometry
provides an enormous advantage over epochal science images due to the
suppression of source confusion in crowded regions and complex backgrounds.
Difference images are generated by IPAC's ZTF real-time production pipeline
and were described by \citet{Masci:18:ZTF}. Section~\ref{sec:submitreqs}
provides a performance benchmark to enable users to estimate the expected
turnaround time following a submission. Instructions on how to retrieve
lightcurve products following submission and processing are described in
\S~\ref{sec:downloadlightcurves}. Recipes for generating publication-quality
lightcurves from the raw output, as well as caveats, warnings, suggestions
for improving their quality, and computing upper limits are described in
\S~\ref{sec:lc}. A method for combining measurements in order to
recover signals below the single-exposure sensitivity limit or place tighter
constraints on non-detections is described in \S~\ref{subsec:coaddmeas}.
A checklist of all user instructions, including a recipe on how to
analyze a ZFPS lightcurve is summarized in \S~\ref{sec:summary}.
Acronyms are defined in Appendix~\ref{acr}.

\clearpage
\section{Overview of the ZFPS}\label{sec:overview}

The ZTF camera generates 64 CCD-quadrant images (also known as readout-channel
images) per exposure, in one of three filters ($g$, $r$, and $i$), and
located on one of the 1338 predefined science Fields that cover the
northern sky visible from Mt. Palomar Observatory. To date, nearly one million
science exposures have been acquired and 44\% of these have products archived
in IRSA. Eventually, all ZTF survey data products will reside in IRSA and be
publicly available. The ZFPS will continue to be operational and integrated
with the ZTF archive following completion of survey operations.

For a given sky position and time span, a forced-photometry lightcurve is
generated for all ZTF camera filters and images according to user
data-access privileges (for example, embargoed ZTF partnership data versus
data that is now publicly accessible). Internally, the sky positions provided
by different users are aggregated according to distinct predefined ZTF Field
and CCD-quadrant locations on the sky.
These are then processed together on the same compute node for maximal
efficiency since the input positions are likely to be from the same
images. This minimizes the I/O load for marshaling large numbers of
difference images onto a given compute node.

\begin{figure*}
\centering
\includegraphics[scale=0.4]{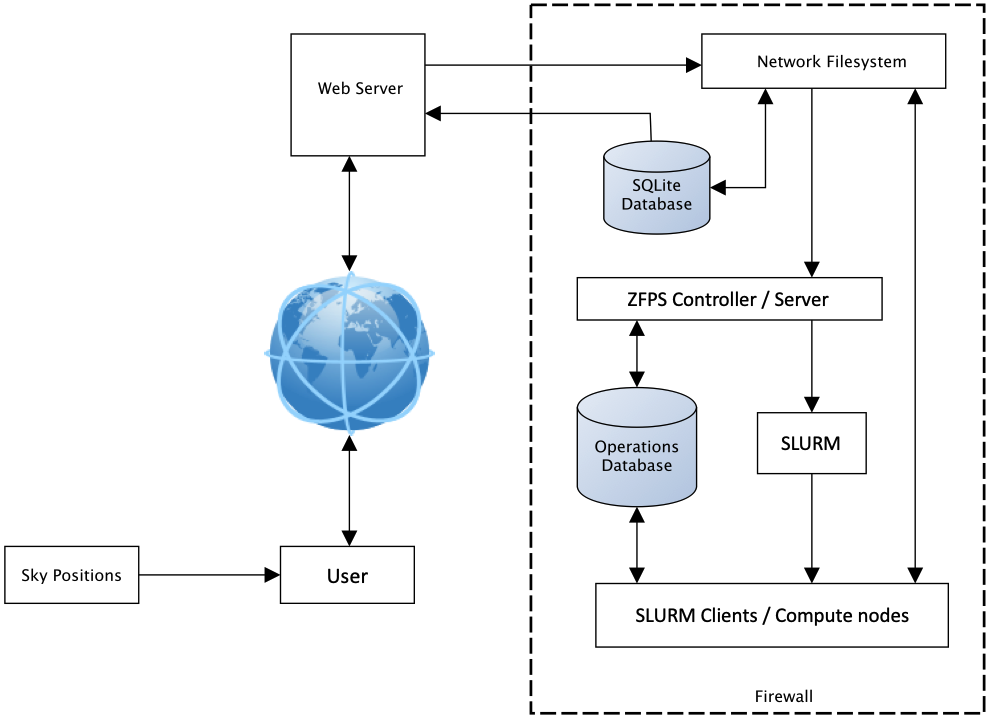}
\caption{ZFPS data flow and system architecture. Major components consist
         of a compute cluster managed by SLURM; an operations database for
         retrieving image inputs and metadata; a SQLite database for storage
         of ZFPS products and a web-server for serving them to users.}
\label{fig:systemflow}
\end{figure*}

Figure~\ref{fig:systemflow} shows the ZFPS data flow along with its major
hardware components. As mentioned earlier, the previous version of this
service processed only one sky position per user submission. The new batch-like
service described here represents a significant improvement in terms of the
number of lightcurves that can be generated and delivered within any given
period. The components and interfaces in Figure~\ref{fig:systemflow} are
described in \S~\ref{sec:methods}.

\clearpage
\section{User Instructions}\label{sec:submission}

Users will first need to register to use the forced-photometry service.
All users who previously registered to use the initial version of the
ZFPS are automatically enrolled and granted the same access privileges
to the new batch-based service. A large portion of the ZTF data is public
and available to everyone; students and post-doctoral scholars are
encouraged to subscribe. New users are invited and can
register by sending an e-mail to {\tt ztf@ipac.caltech.edu}\footnote{This
e-mail address can also be used for helpdesk questions.} requesting
access to the service. It is strongly advised that users who are affiliated
with an academic institution provide their institutional e-mail address.
Individuals should also indicate if they are a member of the ZTF Partnership
or ZTF Project at Caltech. A confidential password (designated {\tt userpass}
below) is assigned to the registered e-mail and sent to the user upon
enrollment.

Prior to submitting requests to the ZFPS (\S~\ref{sec:submitreqs}),
it is advised you check your sky positions for accuracy (at least a few of
them if you have multiple positions), in particular that they reflect what
would be measured directly off IPAC's astrometrically-calibrated images
in the ZTF archive. It is advised that these positions be checked
using epochal images where the signals of your targets are significant
relative to the local noise. Positional biases due to definition of
the astrometric reference system or from uncorrected drifts over time
may exist. We advise that positions be specified in equatorial
coordinates in the ICRF used to define the astrometric system of
Gaia DR1, which is at epoch J2015 \citep{Mignard:Gaia:2016} and is
consistent with the ICRS to $< 1$ mas. This is the astrometric reference
catalog used to astrometrically calibrate all ZTF image and source
catalog products. Furthermore, coordinates defined on the J2000
(or the Earth Mean Equator 2000) system are within 20 mas of the
ICRS \citep{kaplan2005}. This is small enough to ignore when working
with ZTF epochal images since the RMS error in the reconstructed astrometry
per image is typically 60 mas per axis \citep{Masci:18:ZTF}.

Users should also plan on requesting a sufficient number of future and/or
historical epochs after/prior to the epoch or period defining the transient
event or variability behavior of interest. An additional $\sim30$ epochs
is suggested. The purpose is to enable a possible baseline correction
(\S~\ref{subsec:base}) and/or uncertainty rescaling (\S~\ref{subsec:valunc}).
This will also allow the combining of single-epoch measurements in order to go
deeper (\S~\ref{subsec:coaddmeas}). Additional future/historical epochs are
usually unnecessary for continuous reoccurring (a)periodic variables.

\subsection{Submitting Requests}\label{sec:submitreqs}

The official URL of the ZFPS portal is

\begin{quote}
\begin{verbatim}
https://ztfweb.ipac.caltech.edu/batchfp.html
\end{verbatim}
\end{quote}%

\noindent
This webpage will contain periodic updates for users such as the
status of the ZFPS, for example, scheduled maintenance and
downtimes, or updates to the universal ZFPS username or password.

Requests are submitted to the ZFPS using a Python3 script. An example script is
provided in Appendix~\ref{submitcode}. Users can use this template script to
construct their own submission scripts.
The {\tt email} and {\tt userpass} provided to the registered user
following the subscription process must be updated in the script.
The universal {\tt ztffps} account associated with the {\tt auth} keyword in
the HTTP-post-request statement of the script also has a password,
{\tt dontgocrazy!}. This password is different from the {\tt userpass}.
This password should not be updated unless a message was broadcast to
the ZFPS information webpage above.

As a demonstration, the Python script in Appendix~\ref{submitcode} was used
to submit 12,825 sky positions from the ZTF Bright Transient Survey
\citep[BTS;][]{Fremling:20:ZTF,Perley:20:ZTF}, the largest flux-limited
supernova survey to date.\footnote{https://sites.astro.caltech.edu/ztf/bts/bts.php}
As written, the script sent nine separate submissions (via a ``for loop'' at
the bottom), each limited to no more than 1500 sky positions. This is currently
the maximum size of a batch submission and it avoids the following error from
being generated: \hskip 5pt ``{\tt Error:\ 413 Request Entity Too Large}''.

The Python script requires that sky positions be specified in equatorial
coordinates, in decimal format (see \S~\ref{sec:submission} for preferred
astrometric reference system). These coordinates (RA Dec) are supplied in the
hardcoded input file ``List\_of\_RA\_Dec.txt'', with one RA Dec position
per line and a space between the RA and Dec values. The script also requires
the observation start and end Julian Dates (JD) of the survey image epochs.
For reference, the official ZTF-survey start date is
{\tt 2018-03-17 00:00:00.0 UT} ($JD = 2458194.5$~days).

The maximum number of individual (distinct) target positions a user
can have in the system is 15,000.  The user will be notified via e-mail
when this quota is exceeded, not more than once per day. Submitted sky
positions above this limit will wait in a queue on the web server until
the sky positions pending in the system have been processed to a number
below the limit. Furthermore, safeguards are in place to prevent redundant
requests from entering the system too frequently, in order to stave off
possible malicious denial-of-service attacks. Users should refrain from
submitting the same sky positions within a 90-day period. After 90 days,
the same sky positions can be resubmitted and any newly acquired data
will be used to augment the lightcurves.

For reference, the BTS set of lightcurves took $\approx 2.5$~days to compute
in wall-clock terms. This processing occurred and continued during normal
nightly survey operations as data from the observatory was received and
processed at the nominal rate. Turnaround times for lightcurves will be
faster during periods of bad weather, and slower when other data-processing
tasks take priority.

\subsection{Monitoring and Downloading Lightcurves}\label{sec:downloadlightcurves}

Users are notified via e-mail when their jobs are complete and lightcurve
files are ready for download. For the BTS example request
(\S~\ref{sec:submitreqs}), the user was notified via nine separate e-mails,
one for each batch, or in other words, one for each execution of the Python
{\tt requests.post} function in the ``for loop'' of the submission script
(Appendix~\ref{submitcode}). Email notifications occur when all the sky
positions in a specific batch request are completed and could thus be
closed out.

Appendix~\ref{checkcode} contains a script to check the completion of your
submitted requests and if complete, returns a list of the URLs of your
lightcurve file products for download. The returned list is
formatted into {\tt wget ...} lines that can then be executed in
a terminal to download the lightcurve files. Users cannot query the
filenames of lightcurves they do not own because the pathnames include
a unique checksum that is not trivial to decipher. 

The script in Appendix~\ref{checkcode} queries a SQLite database that
contains a 30-day sliding-window history of forced photometry requests.
This database is remade at the top of each hour, and therefore jobs that
finished within the hour will be updated with a lightcurve product filename
and an exit code. The database queries executed by this tool can take
several minutes to execute and return results. Requests older than 30 days
will fall out of the SQLite database and will not be accessible to users
through this tool. However, all lightcurve files and metadata in the
operations database will be stored for the duration of the project,
including those which are recomputed for the same sky positions.

\vspace{0.5cm}
\section{Forced Photometry Methodology}\label{sec:methods}

At the heart of the ZFPS are photometry methods implemented in a C code,
{\tt cforcepsfaper.c}, which utilizes multi-threading for fast
point-spread-function (PSF)-fit photometry, as well as aperture photometry.
The code reads in a time-ordered list of difference images from
the ZTF archive, for all filters ($g$, $r$, $i$), Fields and CCD-quadrants
(or readout-channels) that cover the input sky positions. Only one of
the available filters pertains to a single exposure. The code also reads a
text file containing the ZFPS database request IDs, sky positions, and
corresponding $x,\,y$ positions of the sky locations in each of the
difference-image coordinate frames.
Using this information, the code reads into memory the corresponding
externally upsampled and renormalized PSF-image files (one per difference-image
epoch), and a set of postage-stamp images generated from the difference images.
These stamps have linear dimensions of $25 \times 25$~native pixels and
represent cutouts on the differences images centered on each respective
sky position. Difference images and PSFs are initially generated by the
real-time ZTF pipeline and are described in \citet{Masci:18:ZTF}. 

The number of epochs and sky positions handled by the C code is limited
by the available memory per compute node. The ZTF cluster consists
of 66 compute nodes all of which are managed by the
SLURM software.\footnote{https://slurm.schedmd.com/overview.html}
Each node has at least 16 CPU cores (with two threads admissible per core)
and 125 GB of memory. All nodes are generally available for the ZFPS
when not occupied with either regular nightly real-time processing, scheduled
daily tasks and/or other {\it ad hoc} tasks. The C code is currently
configured to process up to 15,000 epochs and 1,500 sky positions.
The number of sky positions assigned to a given job is limited by the
available memory per target compute node and number of parallel jobs running
on that node. The number of simultaneous threads allowed by an execution
instance of the C code is currently four.

The forced PSF-fit photometry method is simple in that it does not account for
possible contamination by neighboring sources near the target position on
a difference-image stamp. There is no attempt to model and subtract
neighboring point sources or residuals from imperfect subtractions.
Contamination due to source crowding (and/or residuals) is rare, but may be
significant in the galactic plane. Possible overall contamination to
the target signals sought can be accounted for when validating and
rescaling the photometric uncertainties (see~\S~\ref{subsec:valunc}).

\begin{figure*}
\centering
\includegraphics[scale=0.23]{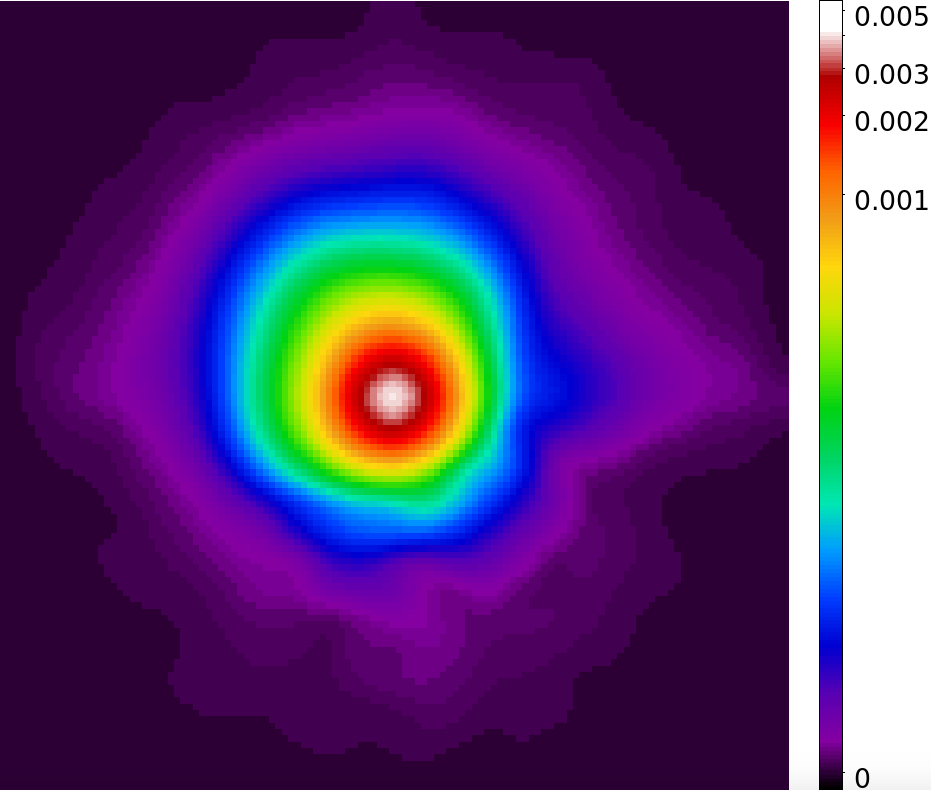}
\includegraphics[scale=0.30]{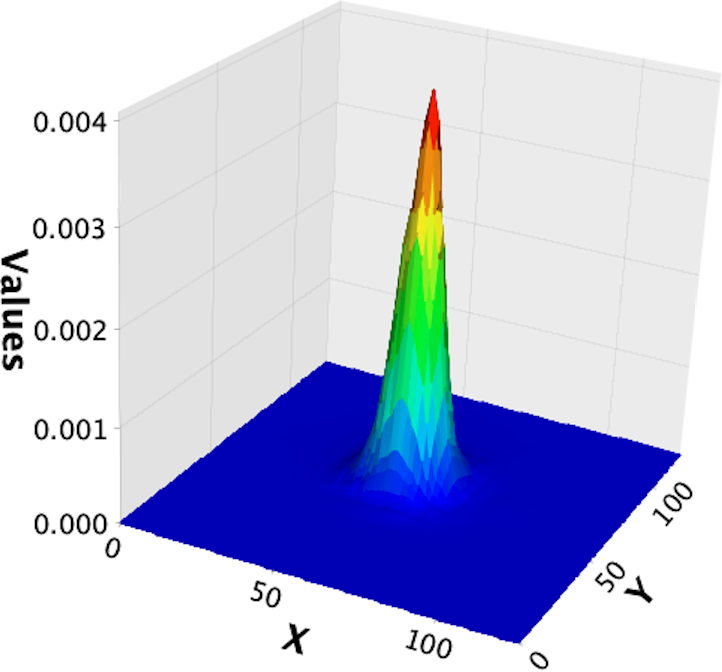}
\caption{Example of an upsampled and recentered PSF. On the {\it left} is a
2-D representation with logarithmic scaling and on the {\it right} is a
3-D representation with linear scaling.}
\label{fig:psf}
\end{figure*}

Each epoch-dependent PSF is first upsampled using Gaussian interpolation
as implemented by the Perl PDL {\tt map} function. This is performed in the
calling Perl program {\tt forcedphotometry\_trim\_cforcepsfaper\_threads.pl}.
PSFs are upsampled by a factor of five per axis. Figure~\ref{fig:psf} shows an
example PSF following upsampling.

The input ($25\,{\rm pixel} \times 25\,{\rm pixel}$) difference-image stamps
are upsampled to match the upsampled PSF-image size, which is
$125 \times 125$~pixels. The upsampling of the difference-image stamps
involves a simple subdivision of each input pixel into $5 \times 5$ sub-pixels
such that the original pixel values are preserved when the sub-pixel values
are summed. The next step is to shift the upsampled difference-image
stamp so that its center pixel coincides with the desired sky-position
to an accuracy that's limited by the upsampled pixel size
(within 1/5 the linear size native pixel or $\sim0.2$ arcsec per axis
for ZTF image data). Difference-image pixels that fall off
the output grid are trimmed and empty pixels are filled with zeroes
following the realignment process.

At this stage, the number of bad pixels in the difference-image stamp
is checked.  If the fraction exceeds 50\% of all pixels, no photometry
is computed and the epoch is flagged with {\tt status = 55}.  If
less than this fraction, the photometry will proceed, but it is
provisionally flagged as a warning with  {\tt status = 56}.
Note that photometric measurements on an epoch-by-epoch basis may be set to
-99999 in the output lightcurve file. This indicates that a measurement
could not be computed (see \S~\ref{sec:exitcodes}).

The difference-image stamps are background-subtracted to account for local
variations in background level (relative to the global zero-level
expected in the input difference images from which the stamps were cut).
The background, $B$, is computed from the median of the values in the
original data prior to upsampling after omitting values within a
radius of 5~pixels from the center. The root-variance of the background
pixel values, $\sigma_{\rm{B}}$, is estimated from half the difference
between the 84th and 16th percentiles (and also used in the
uncertainty calculation described below). This estimate is a robust
alternative to the standard deviation. No photometry is computed if
there are fewer than 100 background pixels. In this case, the epoch is
provisionally flagged with {\tt status = 54}.

\clearpage
The upsampled PSF is converted into a weight map, $W^{\rm{PSF}}_{ij}$, for use
in the PSF-fit photometry step by scaling it by the inverse sum of the
squares of normalized PSF pixel values. That is,

\begin{equation}
S = \sum_{i=1}^{N_s} \sum_{j=1}^{N_s}
{\rm{PSF}}_{ij} \times {\rm{PSF}}_{ij}
\end{equation}

\noindent
and

\begin{equation}
W^{\rm{PSF}}_{ij} = {\rm{PSF}}_{ij}/S \;\; \forall\; i,j
\end{equation}

\noindent
where ${\rm{PSF}}_{ij}$ is the normalized point-spread-function value at
pixel $(i,j)$ and $N_{s} = 125$ pixels.

The PSF-fitted flux, $F^{\rm{PSF}}$, is the summed pixel-by-pixel
product of the weight map and the background-subtracted, upsampled,
recentered difference-image stamp:

\begin{equation}
F^{\rm{PSF}} = C\,\sum_{i=1}^{N_s} \sum_{j=1}^{N_s}
(D_{ij}^{\rm{recen}} - B) W^{\rm{PSF}}_{ij}
\end{equation}

\noindent
where $D_{ij}^{\rm{recen}}$ is the recentered difference-image data at
pixel $(i,j)$, $B$ is the background computed for the stamp image as described
above, $W^{\rm{PSF}}_{ij}$ is the weight-map contribution of the PSF at pixel
$(i,j)$ (Equation 2), and $C$ is a correction factor (currently 0.89897
across all ZTF filters) to account for biases in the estimated flux.
This correction was determined empirically by comparing simulated (truth input)
fluxes to measured fluxes.

The 1-$\sigma$ uncertainty in the PSF-fit flux estimate is given by

\begin{equation}
U^{\rm{PSF}} = \sqrt{\sum_{i=1}^{N_s} \sum_{j=1}^{N_s}
V_{ij} W^{\rm{PSF}}_{ij}W^{\rm{PSF}}_{ij}}
\end{equation}

\begin{equation}
V_{ij} = P_{ij} / G +   \sigma_{\rm{B}}^2 / 25     \;\; \forall\; i,j
\end{equation}

\begin{equation*}
  P_{ij} =\begin{cases}
    (D_{ij}^{\rm{recen}}-B),  &{\text{for}}\; (D_{ij}^{\rm{recen}}-B) > 0 \\
    0  &{\rm{for}}\; (D_{ij}^{\rm{recen}}-B) \le 0 
  \end{cases} \tag{6}
\end{equation*}

\noindent
where $G$ is the effective detector electronic gain, in electrons (e-) per DN
(digital number). The median value for $G$ in each filter across ZTF epochs
is $\sim 6.2$ e-/DN. $\sigma_{\rm{B}}$ is the root variance of the
background (see above).

The signal-to-noise ratio, SNR, is the PSF-fit flux (Equation 3) divided by its
uncertainty (Equation 4):

\addtocounter{equation}{1}
\begin{equation}
{\rm{SNR}} = F^{\rm{PSF}} / U^{\rm{PSF}},
\end{equation}

\noindent
and the reduced chi-squared statistic of the PSF fit, after accounting
for $25 - 1=24$ {\it independent} degrees of freedom following upsampling, is

\begin{equation}
\tilde{\chi}^2 = \frac{1}{24} \sum_{i=1}^{N_s} \sum_{j=1}^{N_s}
[(D_{ij}^{\rm{recen}} - B) - F^{\rm{PSF}} {\rm{PSF}}_{ij}]^2 / V_{ij}
\;\; \forall\; V_{ij} > 0
\end{equation}

Concentric aperture photometry is also computed along with the PSF-fit
photometry and included in the output lightcurve file. The aperture
photometry estimates are generally noisier than the PSF-fitted estimates
and serve as a crude quality check of the PSF-fitted estimates when many
epochal measurements are compared. For a description of aperture photometry
in general and its implementation in a stand-alone tool that is available
for download, see \citet{laher2012}.

\vspace{0.5cm}
\section{Exit and Warning Codes} \label{sec:exitcodes}

Overall exit and warning codes from the ZFPS following execution of each
sky-position request are provided on a per-epoch basis. These are given in
the {\tt procstatus} column of the output lightcurve file.
These codes are defined in Table~\ref{tab:statcodes}. Some of these
codes were described in more detail in \S~\ref{sec:methods}. 
{\tt status = 52} in particular warns the user that the maximum of
10,000 target positions was exceeded. In this case, the user should
submit another request with a later JD range in order to complete
their lightcurves. The 10,000 limit is driven by the local disk space
on a compute node and the number of concurrent forced-photometry jobs
we are allowing per node. 

\begin{deluxetable*}{r|l}
\tablecaption{Status Codes in Lightcurve Files
              ({\tt procstatus} column)\label{tab:statcodes}}
\tablehead{
\colhead{Code} & \colhead{Definition} \\
}
\startdata
0 & Successful execution\\
52 & 10,000 input-file limit reached\\
54 & Insufficient number of background pixels\\
55 & Too many bad pixels\\
56 & Measurement impacted by bad or blank pixels\\
57 & No reference-image-catalog source within 5\arcsec\\
58 & Reference image PSF-catalog does not exist\\
61 & Sky position is off-image or too close to an edge\\
62 & Requested start JD is before official survey start date\\
63 & No records (epochs) returned by database query\\
64 & Catastrophic error\\
65 & Requested end JD is before official survey start date\\
255 & Database or other error\\
\enddata
\end{deluxetable*}

\section{Analyzing a Lightcurve File} \label{sec:lc}

The lightcurve file produced by the ZFPS for a given sky position is a table
formatted in ASCII text. The multi-column data records are space-delimited.
At the beginning of each file is a header containing information about the
request and a description of all the table columns. These lines are preceded
by the {\tt  \#} character. The data portion of the lightcurve file consists
of observational metadata, quality metrics, and forced-photometry measurements
for all epochs at the user's allowed data-access level. Typically, this
lightcurve file is no more than three or four megabytes in size. An example
lightcurve filename is {\tt batchfp\_req0000036025\_lc.txt} where the numeric
string following {\tt \_req} is the unique database request ID.

The data table in a lightcurve file does not represent a standalone
lightcurve that can be immediately plotted. It contains difference-image
forced photometry measurements on the requested input position for the JD
range covering all relevant ZTF Fields, CCD-quadrants, and filters
($g$, $r$, $i$), but only for difference-images that exist in the archive.
Before constructing a lightcurve (\S~\ref{subsec:diffcalmag} and
\S~\ref{subsec:abscalmag}), the contents of the lightcurve file need to be
analyzed such that rows (epochs) corresponding to a specific ZTF Field ID
and filter of interest are stored, including any quality metrics to use
for filtering of bad-quality data (see \S~\ref{subsec:qafilt}).
It's possible that multiple Field IDs
will cover the target position where each Field/CCD-quadrant combination
will have used its own reference image for image-differencing. Each reference
image may have been created from an entirely different set of historical
observation epochs and therefore may introduce a different baseline for
adjusting the resulting differential photometry (see \S~\ref{subsec:base}).
This is primarily due to contamination from either a transient's flux or other
systematic residual in the reference image.

At minimum, the columns (as named in the lightcurve file) for constructing
a forced PSF-fit-photometry differential lightcurve for a given Field,
CCD-quadrant, and filter are: {\tt field}, {\tt ccdid}, {\tt qid},
{\tt filter}, {\tt jd}, {\tt zpdiff}, {\tt forcediffimflux}, and
{\tt forcediffimfluxunc}. The combination of {\tt ccdid} and {\tt qid}
refers to the specific CCD-quadrant of a given {\tt field}. Due to
variations in telescope pointing (and adjustments during the course of the
survey), it is possible for a target position to fall on (or straddle) different
CCD-quadrants of the same {\tt field}. As mentioned above, it is important
that forced-photometry measurements from the same {\tt field}, {\tt ccdid},
{\tt qid}, and {\tt filter} be analyzed (with all necessary corrections
applied; see \S~\ref{subsec:base} and \S~\ref{subsec:valunc}) prior to splicing
them with measurements originating from other combinations of {\tt field},
{\tt ccdid}, and {\tt qid} in the lightcurve file for the same {\tt filter}.

\subsection{Quality Filtering}\label{subsec:qafilt}

Due to the automated nature of ZTF survey operations, a significant fraction
of the data is non-photometric, that is, were acquired through intermittent
cloud cover and/or were contaminated by moonlight, especially reflected from
clouds. Intra-night changes in atmospheric transparency are inevitable and
difficult to avoid or correct in near-realtime. This primarily affects the
real-time photometric calibration (zero-point estimates {\tt zpdiff}).
In particular, spatial non-uniformities in photometric throughput on scales
smaller than a CCD-quadrant ($\lesssim0.85^\circ\,\times\,0.85^\circ$),
however slight, severely impact difference-image quality and measurements
performed thereon.

To account for possible bad photometric measurements, we advise
{\it omitting} epochs that satisfy the three criteria below, based
on a subset of the metrics provided in the lightcurve file.
These were proven to be useful following analyses of large numbers
of Type-Ia SN lightcurves \citep[e.g.,][]{Yao:ZTF:2019}.

\begin{itemize}
\item{Remove epochs with {\tt infobitssci} $\geq 33554432$.
This is a 32-bit integer that encodes the status of
processing and instrumental calibration steps for the science image.
Each condition is assigned a specific bit and not all conditions are
considered ``fatal''. The value 33554432 ($= 2^{25}$)
indicates in general, bad photometric calibration. A stricter
criterion is to remove all epochs with {\tt infobitssci} $> 0$,
but this is likely to remove measurements that could still be
usable, albeit noisier.}
\item{Remove epochs with {\tt scisigpix} $> 25$ DN. This metric is
a robust estimate of the spatial noise-sigma per pixel in the
science image computed via
$0.5~\times~[84\text{th}~-~16\text{th}\; \text{percentiles}]$.}
\item{Remove epochs with ${\tt sciinpseeing} > 4$\arcsec. This
represents the median seeing estimated from the FWHM of point
sources extracted from the science image.}
\end{itemize}

\subsection{Baseline Correction}\label{subsec:base}

\begin{figure*}
\centering
\includegraphics[scale=0.43]{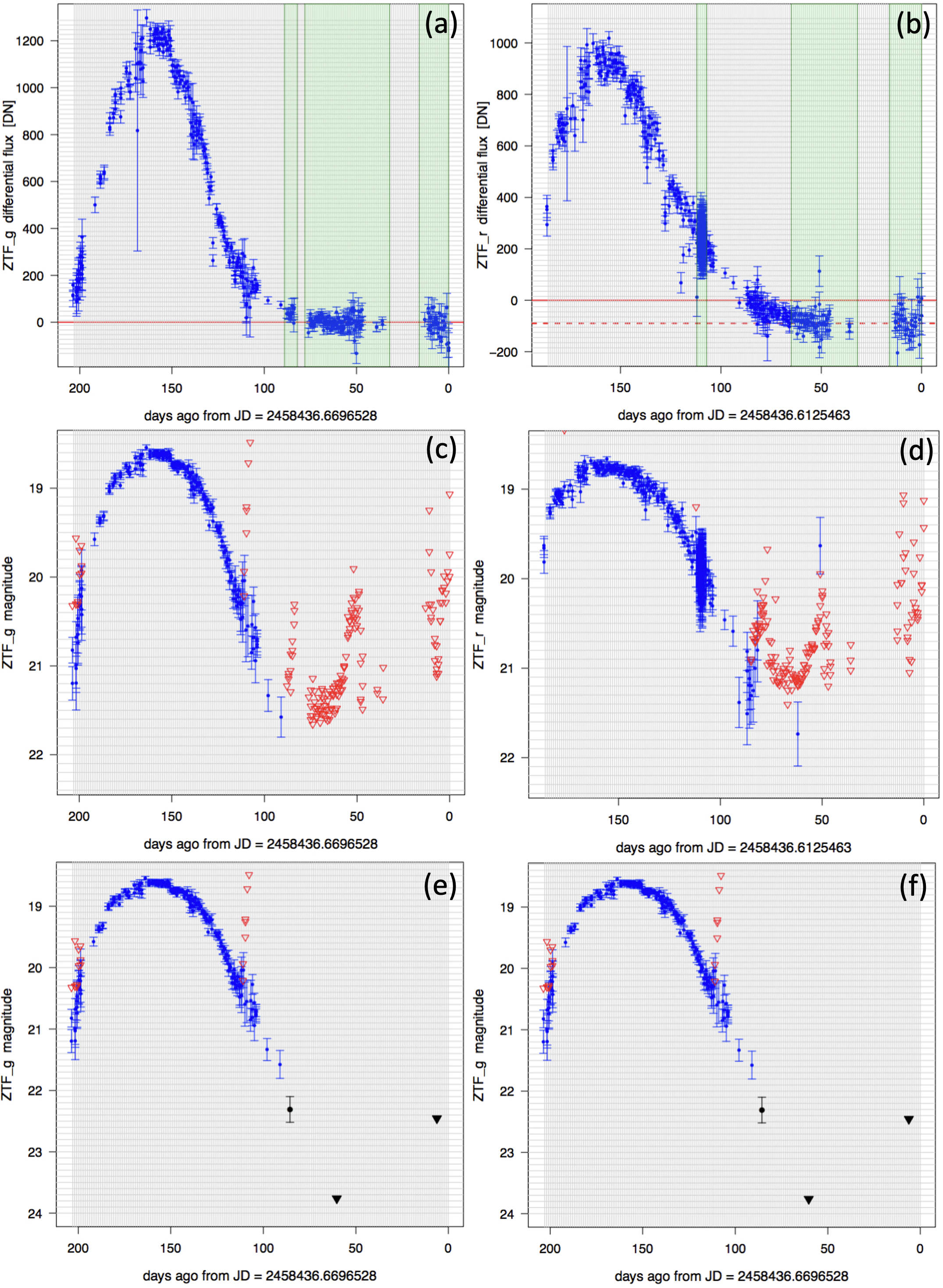} 
\caption{Lightcurves for the Type-I supernova: SN 2018bym. (a) and (b):
instrumental differential-fluxes in the $g$ and $r$ filters respectively
where the green-shaded windows represent timespans where individual
epochs are combined to generate the black symbols in panels (e) and (f); 
(c) and (d): calibrated magnitudes in the $g$ and $r$ filters respectively
where blue symbols are $>3\sigma$ measurements and triangles are $5\sigma$
upper limits; (e) and (f): same as the previous panels but with
combined measurements (black symbols) from a weighted average of the
measurements within the windows shown in (a) and (b).
See \S~\ref{subsec:coaddmeas} for details.} 
\label{lcfigs}
\end{figure*}

For a given {\tt field}, {\tt ccdid}, {\tt qid}, and {\tt filter} in the
lightcurve file and following any filtering of bad data
(\S~\ref{subsec:qafilt}), it is recommended you examine a plot of {\tt forcediffimflux}
[DN]) versus {\tt jd} [day] to determine if there is any residual offset from
the zero-baseline level. This residual offset refers to the level defined
by either historical or future epochs in your lightcurve in flux [DN] space,
where the measured signal (relative to the reference image) is stationary in
time and fluctuating only due to the background noise. For example, if the
reference image used to generate the difference images was contaminated by the
transient’s flux, your baseline will be $< 0$. The reference image may also
be affected by some other systematic during its generation or calibration
(including astrometric registration with the epochal science image), and
therefore the resulting baseline could be $> 0$.

Requesting a special time-range for regenerating a reference-image for the
purpose of ensuring a zero baseline in the differential lightcurve is
unnecessary. You will always need to check and correct for a non-zero
baseline due to possible systematics. This will ensure the best accuracy
following photometric calibration (see \S~\ref{subsec:diffcalmag}).
This correction is not necessary for (a)periodic variables that may fluctuate
about a constant long-term flux-level. The reason is that even if the reference
image was not a true (unbiased) time-average of all the input epochs (at random
lightcurve phases), any residual baseline will need to be retained when
converting the differential fluxes to direct (total) fluxes using the
reference-image source catalog flux (see \S~\ref{subsec:abscalmag}).

The baseline correction needs to use a sufficient number of future and/or
historical flux measurements. We suggest $\ge 30$ epochs. The more the better
since measurements may be affected by stochastic effects, e.g., low
transparency (intermittent cloud cover) and/or moon glow. Before submitting
a ZFPS request, it is recommended you obtain some knowledge of the behavior of
your transient or variability of interest (e.g., timescale) and include a
sufficient number of epochs before and/or after its suspected duration.

Figure~\ref{lcfigs} shows various lightcurves of the Type-I supernova
SN 2018bym \citep{Lunnan:2018}. Flux versus time plots in 
the $g$ and $r$ filters are shown in panels (a) and (b) respectively.
The $r$-filter plot shows a future baseline level
of approximately –90~DN (dashed red line), as inferred from the differential
fluxes at $< 60$~days where the supernova signal reaches quiescence.
In practice, we recommend performing a trimmed-average or median
of the measurements in this ``stationary period'' as guided by a visual
examination of the flux lightcurve. Following an estimate of the baseline
level, you will need to subtract this from all the differential flux
measurements ({\tt forcediffimflux}) before converting them to calibrated
magnitudes or upper limits (see \S~\ref{subsec:diffcalmag}).

\subsection{Validating Flux-Uncertainty Estimates}\label{subsec:valunc}

Despite efforts to ensure the 1-$\sigma$ uncertainties in the PSF-fit fluxes
({\tt forcediffimfluxunc}) are correct, it is advised to check that these are
still plausible, or at least consistent with the photometric repeatability
expected at a given magnitude for the given {\tt filter} and possibly sky
location ({\tt field}), i.e., in the frequentist sense.

The measurement uncertainties are based on propagating a semi-empirical model
of the statistical (random) noise expected in the detector pixels through the
difference- (and reference-) image pipelines. This model does not account for
possible systematics, for example, errors in PSF estimates (such as variation
over a CCD-quadrant), photometric calibration zero-points, flat-fielding
errors, or astrometric calibration residuals, which will impact PSF-placement.

First, we advise checking the distribution of the PSF-fit reduced
$\tilde{\chi}^2$ ({\tt forcediffimchisq}) values for a given {\tt field}
and {\tt filter}, following any quality filtering (\S~\ref{subsec:qafilt}).
This can also be examined
as a function of flux ({\tt forcediffimflux}). The average (or some robust
equivalent) of the {\tt forcediffimchisq} values should be $\approx 1$.
If not, you should multiply the raw {\tt forcediffimfluxunc} values by
$\sqrt{\langle{\tt forcediffimchisq}\rangle}$ where the angled brackets
denote average. Any flux-dependent {\tt forcediffimchisq} could mean the
[e-/DN] gain-factor for estimating the Poisson-noise component is not
properly tuned or the PSF shape and hence flux-scaling is not correct.
Ensuring $\langle{\tt forcediffimchisq}\rangle\approx 1$ only means the pixel
uncertainties are consistent, in a statistical sense, with the PSF-fit
residuals {\it per epoch}. It will not catch and correct for other possible
systematics across the epochs (see below).

Following the {\tt forcediffimchisq} check and possible rescaling, you can
stop here or continue with another check to account for possible temporal
systematics. This can be done by computing a robust RMS (e.g., via
$0.5~\times~[84\text{th}~-~16\text{th}\; \text{percentiles}]$) in the
signal-to-noise (S/N) ratios (i.e.,
${\tt fforcediffimflux} / {\tt fforcediffimfluxunc}$ values) of the
stationary-flux epochs that were used to infer your baseline level
(\S~\ref{subsec:base}). These ratios would be computed following any baseline
correction to the {\tt forcediffimflux} values and any prior scaling
of the {\tt forcediffimfluxunc}. If the uncertainties are plausible, this
RMS should be $\approx 1$. If not, the {\tt forcediffimfluxunc} values can
be multiplied by this RMS value. This will ensure the resulting distribution
in S/N will have zero mean and unit variance. However, beware of epochs with
bad data. This assumes the noise is approximately stationary during the
historical/future epochs and they sample, to first order, the same underlying
background and conditions as in epochs where the transient signal was detected.

The above RMS\{S/N\} check alone will not validate the Poisson-noise component
in epochs when the transient has a high signal-to-noise ratio. A final
(but still somewhat coarse) check for validating specifically the
Poisson-noise-dominated regime can be performed by comparing against plots of
prior-calibrated photometric-repeatability (stack-RMS) versus magnitude
\citep[e.g., see Figure 9 in][]{Masci:18:ZTF}. This analysis will require
you to first perform forced photometry on ``non-variable'' field stars
covering flux ranges of interest, compute stack-RMS dispersions, then
convert the fluxes and RMS values to calibrated magnitudes
(see \S~\ref{subsec:diffcalmag} or \S~\ref{subsec:abscalmag}).
It's important to note that estimates of prior photometric uncertainties
inferred in this manner will be {\tt field} (sky location) dependent.

\subsection{Converting Differential Fluxes to Calibrated Magnitudes}\label{subsec:diffcalmag}

It is assumed the differential fluxes ({\tt forcediffimflux}) have been
corrected for any non-zero baseline (\S~\ref{subsec:base}) and that
flux uncertainties ({\tt forcediffimfluxunc}) have been validated and
possibly rescaled to conform to expectations (see \S~\ref{subsec:valunc}).
If the uncertainties are overestimated or underestimated, the significance
levels and hence derived upper limits will be incorrect. That is, ``5-sigma''
only has meaning if ``sigma'' is correct, and it captures all noise sources
that may have corrupted the measurements.

\clearpage
\noindent
The pseudo-code for obtaining calibrated magnitudes is below. It uses the
following two parameters:

\begin{quote}
\begin{verbatim}
SNT = signal-to-noise threshold for declaring a flux measurement a ``non-detection'' so it
      can be assigned an upper-limit.
SNU = actual signal-to-noise ratio to use when computing a ``SNU-sigma'' upper-limit.
\end{verbatim}
\end{quote}%

\noindent
We recommend SNT~$= 3$ and SNU~$= 5$. For details on why these were picked, see:

\begin{quote}
\begin{verbatim}
http://web.ipac.caltech.edu/staff/fmasci/home/mystats/UpperLimits_FM2011.pdf
\end{verbatim}

\begin{verbatim}
if ( forcediffimflux / forcediffimfluxunc ) > SNT:
    # We have a "confident" detection; compute and plot mag with error bar:
    mag = zpdiff - 2.5 * log10( forcediffimflux )
    sigma_mag = 1.0857 * forcediffimfluxunc / forcediffimflux
else:
    # Compute flux upper limit and plot as arrow or triangle:
    mag = zpdiff - 2.5 * log10( SNU * forcediffimfluxunc )
\end{verbatim}
\end{quote}%

\noindent
Lightcurves of SN 2018bym in the $g$ and $r$ filters converted to magnitudes
using the above construction are shown in panels (c) and (d) of
Figure~\ref{lcfigs}.

\subsection{Total Fluxes for Variable Sources and Conversion to Calibrated Magnitudes}\label{subsec:abscalmag}

Transients are generally defined as sudden events where signal suddenly
appears out of the noise and then disappears. The goal is to analyze the
energy released during the explosive (or perhaps implosive) phase, above or
below some preexisting signal or background. For continuous (a)periodic variables
or transients associated with pre-quiescent sources with detectable signal,
one may want to characterize the {\it total} signal versus time (i.e.,
stationary pre-existing signal plus differential flux component). Here, an
estimate is needed of the source flux from the reference image that was
used to generate the difference-images on which the forced photometry
was performed.

Photometry and quality metrics for the nearest reference-image source
are given in the lightcurve file by columns prefixed with {\tt nearestref...}
These metrics are from the archived reference-image PSF-fit photometry
source catalog. You will first need to threshold on the {\tt dnearestrefsrc}
value to ensure a reference source exists (was extracted) on or close-enough to
your target position. We recommend ${\tt  dnearestrefsrc} < 1$\arcsec.
The {\tt nearestrefchi} and {\tt nearestrefsharp} metrics indicate whether
the reference source is extended or PSF-like. If significantly extended, the
PSF-fit photometry of the nearest reference-image source ({\tt nearestrefmag})
will be erroneous. You will need to ensure the reference-image
extraction is consistent with a point source.

Below are the steps for obtaining a ``total flux'' photometric light curve
in calibrated magnitudes, modulated by variations inferred from differential
photometry.

\begin{enumerate}
\item
Compute the reference source flux and its uncertainty on the same
calibration (zero-point) as the differential flux at each epoch:

\begin{verbatim}
nearestrefflux = 10**[ 0.4 * ( zpdiff - nearestrefmag ) ]
nearestreffluxunc = nearestrefmagunc * nearestrefflux / 1.0857
\end{verbatim}
where {\tt zpdiff}, {\tt nearestrefmag}, and {\tt nearestrefmagunc} are
given in the lightcurve-file data table.

\item
Compute the total flux, its uncertainty, and signal-to-noise ratio at each epoch:

\begin{verbatim}
Flux_tot = forcediffimflux + nearestrefflux
Fluxunc_tot = sqrt[ ( forcediffimfluxunc * forcediffimfluxunc 
                    - nearestreffluxunc * nearestreffluxunc ) ]
SNR_tot = Flux_tot / Fluxunc_tot
\end{verbatim}

The negative sign in the square-root follows from the (implicit) assumption
that the reference source flux was measured from the same reference image used to
generate the ``science – reference'' differential flux. Hence, the noise is
anti-correlated. If this is not true or if the argument in the square-root is
$\le 0$ for whatever reason, we suggest adding the variances instead.
This will be a conservative estimate.

\item
Using the outputs from item 2 and the SNT, SNU parameters defined in
\S~\ref{subsec:diffcalmag}, here is the pseudo-code for generating a
``total-flux'' photometric lightcurve:

\begin{quote}
\begin{verbatim}
if SNR_tot > SNT:
    # We have a "confident" detection; compute and plot mag with error bar:
    mag = zpdiff - 2.5 * log10( Flux_tot )
    sigma_mag = 1.0857 / SNR_tot
else:
    # Compute flux upper limit and plot as arrow or triangle:
    mag = zpdiff - 2.5 * log10( SNU * Fluxunc_tot )
\end{verbatim}
\end{quote}%

This logic reduces to the differential-lightcurve generation logic in
\S~\ref{subsec:diffcalmag} when
${\tt nearestrefflux} = 0$ and ${\tt Fluxunc\_tot} = {\tt forcediffimfluxunc}$.

\end{enumerate}

The above construction to compute calibrated magnitudes assumes the
source of interest has zero color in the Pan-STARRS1 photometric system
($g_{\rm{PS1}} - r_{\rm{PS1}} = 0$ for $g$- and $r$-flux measurements;
$r_{\rm{PS1}} - i_{\rm{PS1}} = 0$ for $i$-flux measurements).
If you have prior information on the color or spectrum of your source in
these bandpasses across epochs, the magnitude computations above can include
the additive term: $+ {\tt clrcoeff} * (g_{\rm{PS1}} - r_{\rm{PS1}}$) for
$g$ and $r$ measurements, or $+ {\tt clrcoeff} * (r_{\rm{PS1}} - i_{\rm{PS1}}$)
for $i$-filter measurements, where {\tt clrcoeff} is a data column
in the lightcurve file.

\subsection{Going Deeper: combining single-epoch measurements}\label{subsec:coaddmeas}

To make your lightcurve measurements more statistically significant and/or
make upper-limit estimates tighter, you can attempt to combine the
flux measurements within carefully selected time-windows using some optimal
method. Before proceeding, it is assumed your single-epoch differential
fluxes have been corrected for any non-zero baseline (in the case of
transients; see \S~\ref{subsec:base}) and uncertainties have been
validated and rescaled if necessary (see \S~\ref{subsec:valunc}).

In order to improve SNR from combining measurements, the important
assumption is that the individual measurement errors be uncorrelated.
This is generally true when the individual measurements on a target's
position are background (hence photon and/or read noise) dominated.
In the case of image differencing however, systematics from imperfect
subtractions from e.g., inaccurate PSF-matching, astrometric registration,
and/or photometric-gain matching will dominate and persist beyond some
limit where $\sqrt{N}$ improvements stop.
This limit is difficult to determine without use of prior information
or constraints on your transient's temporal behavior. For the most part
however, the ZTF single epoch difference-image measurements are photon
and/or read-noise dominated and some improvement in SNR is always possible
by combining up to several tens of epochs before systematics will
limit further improvement.

Before combining measurements, you will first need to place the fluxes on
the same (fiducial) photometric
zeropoint (call this $\overline{ZP}$). You can pick any value for
$\overline{ZP}$. A value that’s close to the epoch-dependent zeropoints
(${\tt zpdiff}_i$ values, where $i$ is an index over the values) will
suffice. You will then rescale the corrected fluxes ({\tt forceddiffimflux})
and uncertainties ({\tt forceddiffimfluxunc}) as follows:

\begin{equation}
{\tt forceddiffimflux}_i^{\rm{New}}= {\tt forceddiffimflux}_i \times 10^{0.4(\overline{ZP} - {\tt zpdiff}_i)}
\end{equation}

\begin{equation}
{\tt forceddiffimfluxunc}_i^{\rm{New}}= {\tt forceddiffimfluxunc}_i \times 10^{0.4(\overline{ZP} - {\tt zpdiff}_i)}
\end{equation}

\noindent
One method for combining the measurements is to assume the underlying source
signal is stationary within a time-window and collapse the $N$ rescaled
single-epoch fluxes (${\tt forceddiffimflux}_i^{\rm{New}}$) therein using an
inverse-variance weighted average:

\begin{equation}
{\tt flux}^{\rm{New}} = \frac{\sum^N_i{w_i {\tt forceddiffimflux}_i^{\rm{New}}} } {
  \sum^N_i{w_i} }
\end{equation}

\noindent
where

\begin{equation}
w_i = \frac{1} {[{\tt forceddiffimfluxunc}_i^{\rm{New}}]^2}
\end{equation}

\noindent
The uncertainty in ${\tt flux}^{\rm{New}}$ is:

\begin{equation}
{\tt fluxunc}^{\rm{New}} = \frac{1}{\sqrt{\sum^N_i{w_i}}} \label{eq:fluxuncnew}
\end{equation}

\noindent
Equation~\ref{eq:fluxuncnew} is effectively
$\approx {\tt forceddiffimfluxunc}^{\rm{New}}/\sqrt{N}$ if one assumes
the $i=1...N$ windowed single-epoch flux measurements have approximately the
same uncertainty equal to some constant value or RMS. Therefore, the
improvement in signal-to-noise ratio assuming uncorrelated measurements is no
more than a factor of $\approx\sqrt{N}$ (however beware of
systematic errors; see above). The important assumption is that the
underlying source signal is constant within the window. It may appear constant
within measurement error, but when many measurements are available, you can
attempt to bin them in various ways to tease out possible hidden trends in
the signal. A moving (weighted) average may also work, using a properly tuned
window size. There is also a large collection of methods on local-polynomial
regression fitting \citep[e.g.,][]{Buja:1989,Ledolter:2008}.

You can also attempt to collapse the measurements within windows by fitting a
prior model of flux versus time, i.e., if you have prior (or contextual)
knowledge from other data sources. The advantages of combining
measurements in source-space across epochs as opposed to co-adding entire
images in time-ordered slices are: (i) speed and (ii) greater flexibility
when weighting and combining the data.

Having computed the window-collapsed fluxes (${\tt flux}^{\rm{New}}$),
you can now convert them to calibrated magnitudes with upper limits using the
formalism in \S~\ref{subsec:diffcalmag} (for transients) or
\S~\ref{subsec:abscalmag} (for ``total'' photometry on re-occurring
variable sources). The only difference is that you will
need to use the $\overline{ZP}$ value assumed when rescaling the input fluxes
(see above). Panels (a) and (b) in Figure~\ref{lcfigs} show example windows
(shaded green) where the single-epoch flux measurements were combined using
a weighted average (Equations 11 and 13). The resulting calibrated measurements 
in the $g$ and $r$ filters with upper limits are shown in panels (e) and
(f) of Figure~\ref{lcfigs}.

\clearpage
\section{Summary and Checklist}\label{sec:summary} 

We have described the recently upgraded ZFPS, which now supports
the submission of batch requests directly from a Python script.
Instructions and guidelines for analyzing and generating
photometrically-calibrated, publication-quality lightcurves
were also presented. Below is a checklist with references to
sections containing further details.

\begin{enumerate}
\item{If you are interested in using the service, please register by sending
      an e-mail to {\tt ztf@ipac.caltech.edu}. Include your affiliation with
      the ZTF Project (if any) as outlined in \S~\ref{sec:submission}.
      Follow the advisories in \S~\ref{sec:submission} prior to submitting
      your targets.}
\item{Consult the ZFPS information page: {\tt https://ztfweb.ipac.caltech.edu/batchfp.html}
      for important updates relating to the service (\S~\ref{sec:submitreqs}).}
\item{Prepare your input file list of target RA Dec positions, pick the JD
      range of interest for your lightcurves, and follow the steps in
      Appendix~\ref{submitcode} to submit your request. Also see details in
      \S~\ref{sec:submitreqs}.}      
\item{You will be notified by email when your jobs are complete and lightcurve
      files are ready for
      download. You can monitor the status of your request by modifying and
      uploading the example script in Appendix~\ref{checkcode}. Follow the
      steps therein to retrieve your lightcurve files when the script returns
      the {\tt wget ...} lines. Also see details in
      \S~\ref{sec:downloadlightcurves}.}
\item{To extract a lightcurve from a downloaded lightcurve file for a target 
      position and given {\tt filter} (= $g$, $r$, or $i$), ensure that only
      rows with the same values of {\tt field}, {\tt ccdid}, {\tt qid}, and
      {\tt filter} are extracted, as discussed in \S~\ref{sec:lc}. At minimum,
      store the following corresponding quantities and quality metrics:
      {\tt jd}, {\tt zpdiff}, {\tt forcediffimflux}, {\tt forcediffimfluxunc},
      {\tt infobitssci}, {\tt scisigpix}, and {\tt sciinpseeing}.}
\item{Remove bad-quality measurements from the records stored in the
      previous step by thresholding on {\tt infobitssci}, {\tt scisigpix},
      and {\tt sciinpseeing} using the suggested thresholds in
      \S~\ref{subsec:qafilt}.}
\item{Examine a plot of {\tt forcediffimflux} vs {\tt jd} and perform a
      baseline correction if necessary as described in \S~\ref{subsec:base}.
      A baseline correction will not be necessary for the differential-flux
      lightcurves of (a)periodic variables for which you plan to convert to
      ``total flux'' lightcurves later (\S~\ref{subsec:abscalmag}; see also
      step 10 below).}
\item{Check whether the flux uncertainties ({\tt forcediffimfluxunc}) are
      plausible using the tests suggested in \S~\ref{subsec:valunc}. If not,
      compute correction factors and correct the {\tt forcediffimfluxunc}
      values.}
\item{If your lightcurve corresponds
      to transient behaviour over some finite timespan, convert the
      differential fluxes to calibrated magnitudes or upper limits following
      the recipe in \S~\ref{subsec:diffcalmag}.}
\item{If you are interested in recovering the {\it total} signal versus time for
      a lightcurve and wish to convert total instrumental fluxes to
      calibrated magnitudes, follow the recipe in \S~\ref{subsec:abscalmag}.}
\item{If you are interested in combining single-epoch measurements in order
      to improve SNR or place tighter upper limits on non-detections, follow
      the suggestions in \S~\ref{subsec:coaddmeas}.}
\end{enumerate}


\vspace{0.6cm}
\noindent
Based on observations obtained with the Samuel Oschin Telescope 48-inch and
the 60-inch Telescope at the Palomar Observatory as part of the Zwicky
Transient Facility project. ZTF is supported by the National Science
Foundation under Grants No. AST-1440341 and AST-2034437 and a collaboration
including current partners Caltech, IPAC, the Weizmann Institute of Science,
the Oskar Klein Center at Stockholm University, the University of Maryland,
Deutsches Elektronen-Synchrotron and Humboldt University, the TANGO
Consortium of Taiwan, the University of Wisconsin at Milwaukee, Trinity
College Dublin, Lawrence Livermore National Laboratories, IN2P3,
University of Warwick, Ruhr University Bochum, Northwestern University and
former partners the University of Washington, Los Alamos National
Laboratories, and Lawrence Berkeley National Laboratories. Operations are
conducted by COO, IPAC, and UW. The ZTF forced-photometry service was funded
under the Heising-Simons Foundation grant \#12540303 (PI: M. J. Graham).


\clearpage
\facilities{PO:1.2m}

\software{Python3~(\url{https://www.python.org}),
          Perl5~(\url{https://www.perl.org}),
          PDL~(\url{https://pdl.perl.org}),
          GCC~(\url{https://gcc.gnu.org}),
          CFITSIO~(\url{https://heasarc.gsfc.nasa.gov/fitsio/})}

\vspace{0.5cm}
\bibliographystyle{aasjournal}
\bibliography{references}

\appendix

\section{Submission Script Example}\label{submitcode}

Below is an example script for submitting requests to the ZFPS, as
described in \S~\ref{sec:submitreqs}. Instructions are as follows:

\begin{enumerate}
\item{Ensure you have a version of Python3 installed (preferably
      version $\geq 3.7$). Also ensure the {\tt requests} and
      {\tt json} Python packages are installed.}
\item{Copy and paste the lines that fall between the {\tt ``$>>>>>>>>>>$''}
      lines into a file named e.g., zfps\_submit.py.}
\item{Make this file an executable script. On a Linux/Unix operating system,
      this is the command: ``chmod +x zfps\_submit.py''.}
\item{Update the JD range of interest. These are the {\tt jds}
      and {\tt jde} values in the script.}
\item{Supply your {\tt email} and {\tt userpass} where indicated inside
      the script.}
\item{Prepare a file containing a list of RA Dec positions in 
      decimal degrees; one per line where the RA Dec are space-separated.
      This file is named ``List\_of\_RA\_Dec.txt'' in the script.}
\item{Execute the script to submit your batch request. On a Linux/Unix
      operating system, you would execute this at the command-prompt as 
      ``./zfps\_submit.py''.}
\end{enumerate}

\vspace{0.1cm}
\begin{quote}
\begin{verbatim}
>>>>>>>>>>>>>>>>>>>>>>>>>>>>>>>>>>>>>>>>>>>>>>>>>>>>>>>>>>>>>>>>>>>
#!/usr/bin/env python3

import requests
import json

# Script name: zfps_submit.py 

def submit_post(ra_list,dec_list):

    ra = json.dumps(ra_list)
    print(ra)
    dec = json.dumps(dec_list)
    print(dec)

    jds = 2458216.1234            # start JD for all input target positions.
    jdstart = json.dumps(jds)
    print(jdstart)

    jde = 2458450.0253            # end JD for all input target positions.
    jdend = json.dumps(jde)
    print(jdend)

    email = 'joeastronomer@caltech.edu'       # email you subscribed with.
    userpass = 'asdf123'                      # password that was issued to you.

    payload = {'ra': ra, 'dec': dec, 
               'jdstart': jdstart, 'jdend': jdend,
               'email': email, 'userpass': userpass}

    # fixed IP address/URL where requests are submitted:
    url = 'https://ztfweb.ipac.caltech.edu/cgi-bin/batchfp.py/submit'

    r = requests.post(url,auth=('ztffps', 'dontgocrazy!'), data=payload)
    print("Status_code=",r.status_code)

#--------------------------------------------------
# Main calling program. Ensure "List_of_RA_Dec.txt"
# contains your RA Dec positions.

with open('List_of_RA_Dec.txt') as f: 
    lines = f.readlines()
f.close()

print("Number of (ra,dec) pairs =", len(lines))

ralist = []
declist = []
i = 0
for line in lines:
    x = line.split()
    radbl = float(x[0])
    decdbl = float(x[1])

    raval = float('%.7f'%(radbl))
    decval = float('%.7f'%(decdbl))

    ralist.append(raval)
    declist.append(decval)

    i = i + 1
    rem = i % 1500    # Limit submission to 1500 sky positions.

    if rem == 0:
        submit_post(ralist,declist)
        ralist = []
        declist = []

if len(ralist) > 0:
    submit_post(ralist,declist)
   
exit(0)
>>>>>>>>>>>>>>>>>>>>>>>>>>>>>>>>>>>>>>>>>>>>>>>>>>>>>>>>>>>>>>>>>>>
\end{verbatim}
\end{quote}%

\clearpage
\section{Job-Status Checking Script}\label{checkcode}

Below is a script to check the completion of your submitted requests and
if complete, returns a list of the URLs of your lightcurve file products
for download as described in \S~\ref{sec:downloadlightcurves}. 
Instructions are as follows:

\begin{enumerate}
\item{Ensure you have a version of Python3 installed (preferably 
      version $\geq 3.7$). Also ensure the {\tt requests} and
      {\tt re} Python packages are installed.}
\item{Copy and paste the lines that fall between the {\tt ``$>>>>>>>>>>$''}
      lines into a file named e.g., check\_status.py.}
\item{Make this file an executable script. On a Linux/Unix operating system,
      this is the command: ``chmod +x check\_status.py''.}
\item{Supply your {\tt email} and {\tt userpass} where indicated inside
      the script.}
\item{If you are interested in the status of {\it recently} subitted jobs,
      ensure {\tt 'option':} in the script is set to {\tt 'All recent jobs'}
      (as currently indicated). If you are interested in checking for
      pending jobs, ensure {\tt 'option':} is set to {\tt 'Pending jobs'}.}
\item{Execute the script. On a Linux/Unix operating system, you would
      execute this at the command-prompt as ``./check\_status.py''.}
\end{enumerate}

\noindent
Following execution, the script will either return a message such
as {\tt Jobs either queued ...}, or if complete, an actual listing
of the URLs of the lightcurve files formatted into {\tt wget ...} lines.
These lines can be pasted and executed in a terminal window to
download the lightcurve files. Below is an example output following
completion of a request consisting of four sky positions. The ``$\backslash$''
are not part of the output; they are used here for splitting the long
output lines.

\begin{verbatim}
./check_status.py 
Script executed normally and queried the ZTF Batch Forced Photometry database.

wget --http-user=ztffps --http-passwd=dontgocrazy! -O  batchfp_req0000043161_lc.txt \
     "https://ztfweb.ipac.caltech.edu/ztf/ops/forcedphot/lc/batchfp/br000001-000100/32/\
     3f30e9eb728027c210739d20151911be/batchfp_req0000043161_lc.txt"
wget --http-user=ztffps --http-passwd=dontgocrazy! -O  batchfp_req0000043162_lc.txt \
     "https://ztfweb.ipac.caltech.edu/ztf/ops/forcedphot/lc/batchfp/br000001-000100/32/\
     3f30e9eb728027c210739d20151911be/batchfp_req0000043162_lc.txt"
wget --http-user=ztffps --http-passwd=dontgocrazy! -O  batchfp_req0000043163_lc.txt \
     "https://ztfweb.ipac.caltech.edu/ztf/ops/forcedphot/lc/batchfp/br000001-000100/33/\
     dc0d134536adc64b54df088eae99bd78/batchfp_req0000043163_lc.txt"
wget --http-user=ztffps --http-passwd=dontgocrazy! -O  batchfp_req0000043164_lc.txt \
     "https://ztfweb.ipac.caltech.edu/ztf/ops/forcedphot/lc/batchfp/br000001-000100/33/\
     dc0d134536adc64b54df088eae99bd78/batchfp_req0000043164_lc.txt"
\end{verbatim}

\noindent
Alternatively, the URL embedded in the script below, that is:

\begin{quote}
\begin{verbatim}
https://ztfweb.ipac.caltech.edu/cgi-bin/getBatchForcedPhotometryRequests.cgi
\end{verbatim}
\end{quote}%

\noindent
can be pasted into a browser in which case a webform will be displayed
asking for your log-in details.

\clearpage
\begin{quote}
\begin{verbatim}
>>>>>>>>>>>>>>>>>>>>>>>>>>>>>>>>>>>>>>>>>>>>>>>>>>>>>>>>>>>>>>>>>>>
#!/usr/bin/env python3

import re
import requests

# Script name: check_status.py

settings = {'email': 'joeastronomer@caltech.edu','userpass': 'asdf123',
            'option': 'All recent jobs', 'action': 'Query Database'}

r = requests.get('https://ztfweb.ipac.caltech.edu/cgi-bin/' +\
                 'getBatchForcedPhotometryRequests.cgi',
                 auth=('ztffps', 'dontgocrazy!'),params=settings)
#print(r.text)

if r.status_code == 200:
   print("Script executed normally and queried the ZTF Batch " +\
         "Forced Photometry database.\n")

   wget_prefix = 'wget --http-user=ztffps --http-passwd=dontgocrazy! -O '
   wget_url = 'https://ztfweb.ipac.caltech.edu'
   wget_suffix = '"'
   lightcurves = re.findall(r'/ztf/ops.+?lc.txt\b',r.text)

   if lightcurves is not None:
      for lc in lightcurves:
         p = re.match(r'.+/(.+)', lc)
         fileonly = p.group(1)
         print(wget_prefix + " " + fileonly + " \"" + wget_url + lc +\
               wget_suffix)
else:
   print("Status_code=",r.status_code,"; Jobs either queued or" +\
         "abnormal execution.")
>>>>>>>>>>>>>>>>>>>>>>>>>>>>>>>>>>>>>>>>>>>>>>>>>>>>>>>>>>>>>>>>>>>
\end{verbatim}
\end{quote}%

\clearpage
\section{Acronyms}\label{acr}

\begin{table}[!ht]
{\small
{
\renewcommand{\arraystretch}{1.0}
\begin{tabular}{@{}ll}
AGN   & Active Galactic Nuclei\\
ASCII & American Standard Code for Information Interchange\\
BTS   & Bright Transient Survey\\
CCD   & Charge Coupled Device\\
COO   & Caltech Optical Observatories\\
CPU   & Central Processing Unit\\
Dec   & Declination\\
DN    & Digital Number\\
DR    & Data Release\\
FITS  & Flexible Image Transport System\\
FWHM  & Full Width at Half Maximum\\
HTTP  & Hypertext Transfer Protocol\\
ICRF  & International Celestial Reference Frame\\
ICRS  & International Celestial Reference System\\
ID    & Identifier\\
IN2P3 & L'Institut national de physique nucl\'eaire et de physique des particules\\
IO    & Input/Output\\
IRSA  & NASA/IPAC Infrared Science Archive\\
JD    & Julian Date\\
mas   & Milli-arcseconds\\
NASA  & National Aeronautics and Space Administration\\
PDL   & Perl Data Language\\
PI    & Principal Investigator\\
PS1   & Pan-STARRS1\\
PSF   & Point Spread Function\\
qid   & Quadrant Identifier\\
RA    & Right-Ascension\\
RMS   & Root Mean Square deviation or error\\
SLURM & Simple Linux Utility for Resource Management\\
SN    & Supernova\\
SNR   & Signal-to-Noise Ratio\\
SNT   & Signal-to-Noise Threshold (for declaring significant detection)\\
SNU   & Signal-to-Noise Upper Limit (for assigning upper limit)\\
SQL   & Structured Query Language\\
URL   & Uniform Resource Locator\\
UT    & Universal Time\\
UW    & University of Washington\\
ZFPS  & ZTF Forced Photometry Service\\
ZP    & ZeroPoint (photometric-zeropoint)\\
ZTF   & Zwicky Transient Facility\\
\end{tabular}
}
}
\end{table}

\end{document}